\documentclass[reprint,amsmath,amssymb,aps,prl,superscriptaddress]{revtex4-1}

\usepackage{graphicx,textcomp,mathtools,amsmath,booktabs}

\begin{document}

\title{Determination of fracture toughness of thin-film amorphous silicon using
  spiral crack structures}

\author{Torsten Bronger}
\affiliation{Forschungszentrum J\"ulich GmbH, Institut f\"ur Energie und
  Klimaforschung (IEK-5), J\"ulich, 52425 (Germany)}

\begin{abstract}
  We prepared thin layers of amorphous silicon by deposition of a liquid-phase
  polysilane precurser on glass substrate.  Raman scattering provides evidence
  for residual tensile stress in the silicon, which is evaluated
  quantitatively.  Under treatment with hydrofluoric acid, this stress leads to
  spiral cracks in the silicon.  We explain the process of crack formation and
  examine this phenomenon both analytically and numerically, the latter with
  the finite element method (FEM)\@.  The FEM yields the geometry correction
  factor for such spiral cracks in terms of the Griffith criterion.  This
  allows for the first time the determination of fracture toughness of
  amorphous silicon, which is greatly enhanced in comparison with crystalline
  silicon.
\end{abstract}

\maketitle

\section{Introduction}

Semiconductor layers made from liquid-phase precursors have gained interest in
recent years.  On the one hand, they promise low production costs, and on the
other hand, device quality could be increased significantly.  In particular, as
recently reported, the efficiency of solar cells made of liquid silicon
precursors reaches 3.5\,\%. \cite{bronger2014solution}

One challenge of this technique is the forming of a low-stress layer.  The
cause of the residual stress is largely unknown.  Possible candidates are the
high temperature change as well as the chemical processes during the drying.
Stress in thin layers deposited on substrates is a topic of interest due to
potential delamination of the layer.  Such delamination usually renders the
sample or the device unusable.  It is caused by interfacial cracking between
layer and substrate in case the stress in the layer is large enough.  Moreover,
stress may induce the formation of voids in the material, and it has been shown
to cause defects in amorphous silicon (\mbox{a-Si:H})~\cite{stutzmann1985role}.
For amorphous silicon prepared by plasma-enhanced chemical vapour deposition,
the stress is well-controlled to a point that it usually imposes no problem for
the resulting device.  However, for other material systems like
micro-crystalline silicon, or other fabrication techniques like the deposition
from liquid-phase precursors as used in our work, it may limit the achievable
thickness or device quality.  \cite{bronger2014solution}

For characterising stress in silicon, Raman spectroscopy is a well-established
method \cite{anastassakis1970effect}, also in the micro-Raman
variant~\cite{de1996micro}.  There exists a linear relationship between
in-plane stress and the shift of the Si--Si peaks.
\cite{anastassakis1990piezo} In case of a single crystal, this shift may occur
in conjunction with a peak split for crystalline material.
\cite{anastassakis1990piezo}

Another way for stress to become visible is the generation of cracks.  A small
initial weakness in the layer may be the origin of a crack propagating and thus
relaxing the stressed layer.  Theoretically, this phenomenon can be analysed
both analythically (important breakthroughs were the Griffith theory founded in
\cite{griffith1920vi} and the introduction of the stress intensity factor in
\cite{irwin1957analysis}), and numerically by the finite element method FEM
\cite{chan1970finite} and the boundary element method BEM
\cite{blandford1981two}.

In this work, we will describe spiral crack formation in etched amorphous
silicon layers prepared from a liquid-phase precursor.  We will present a model
for the development of the cracks.  Finally, we will use this model for
analysing the observations with an analogy to Griffith's theory as well as with
extensive FEM simulations, yielding the fracture toughness of the silicon
material.

\section{Experiments}

\begin{figure}
  \centering
  \includegraphics[width=0.9\linewidth]{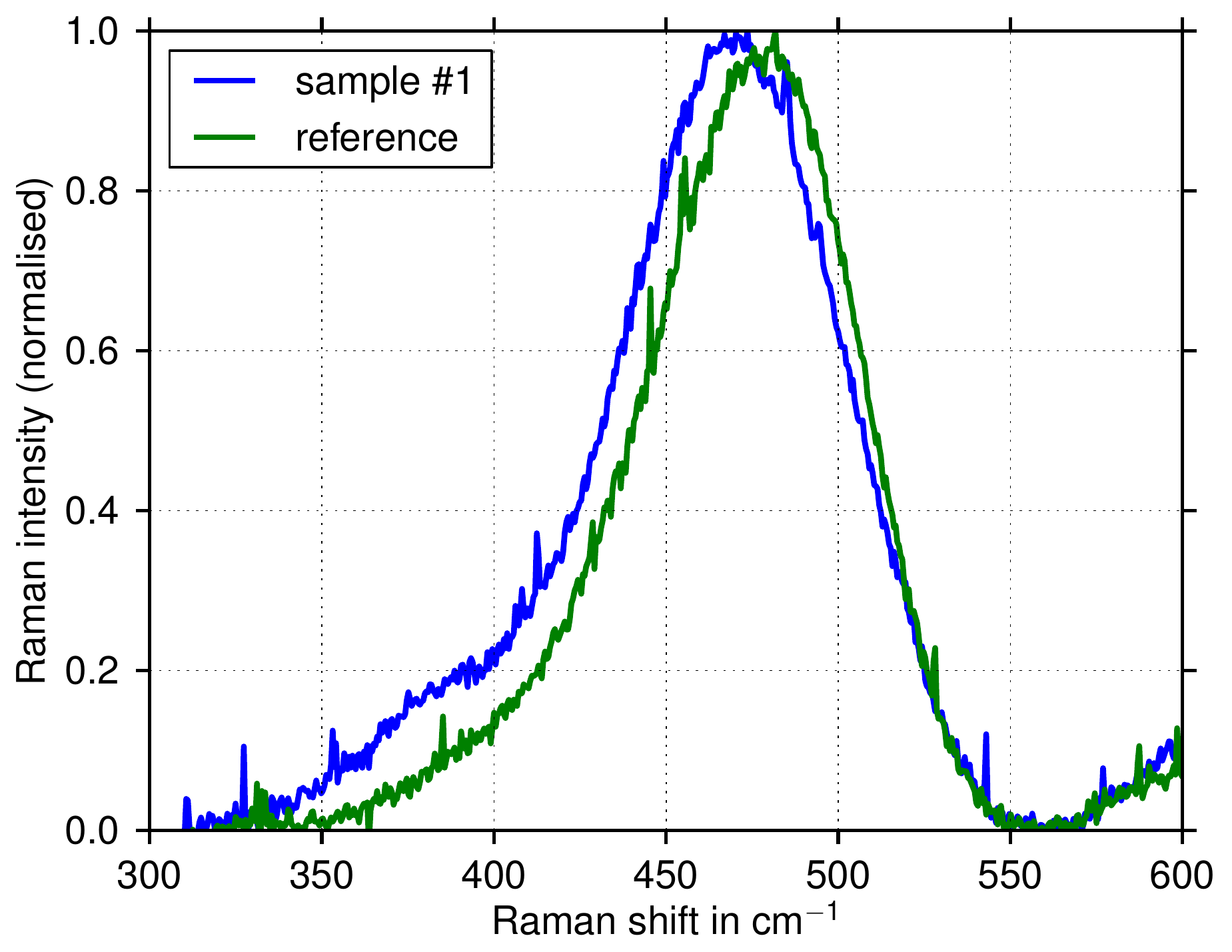}
  \caption{Raman spectra of the stressed sample \#1 and a PECVD
    \mbox{a-Si:H} reference sample with 700\,nm thickness.  The laser excitation
    wavelength is 488\,nm.}
  \label{fig:raman}
\end{figure}

\begin{figure}
  \centering
  \includegraphics[width=0.9\linewidth]{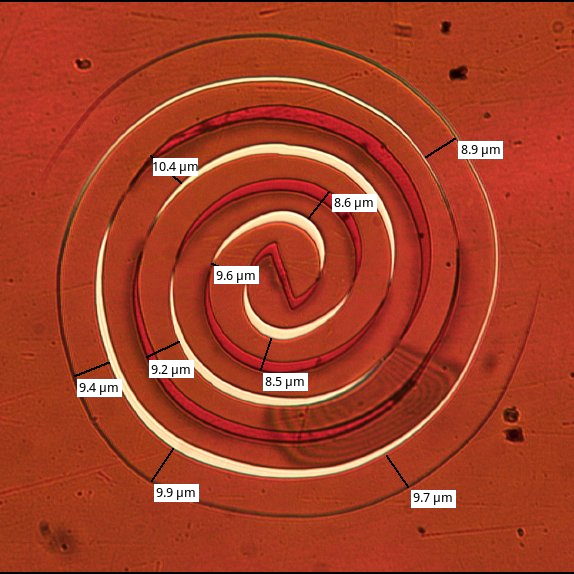}
  \caption{Transmitted-light microscope image of a crack structure in an
    \mbox{a-Si:H} layer.  Yellow areas denote lack of layer material, i.\,e.\
    only the substrate is visible.  Two material layers on top of each other
    appear in reddish colour.  At various places, the stripe width is measured,
    and the resulting value is shown.}
  \label{fig:crack}
\end{figure}

\begin{table}
  \centering
  \begin{tabular}{@{}ccc@{}}
    \toprule
    sample \# & Raman peak shift $\Delta\omega$  & stripe width $w$ \\
             & in cm$^{-1}$ & in \textmu m \\
             \midrule
             1 & $-9.1$ & \phantom07.5 \\          
             2 & $-9.0$ & \phantom09.5 \\          
             3 & $-7.7$ & 10.0\\   
             4 & $-6.5$ & \phantom08.0\\   
    \bottomrule
  \end{tabular}
  \caption{Measured shift if the \mbox{a-Si:H} Raman peak and spiral stripe
    width for \mbox{a-Si:H} samples.}
  \label{tab:experimental-results}
\end{table}

The samples are single layers of amorphous silicon, made from soluble
polysilane precursor, deposited on glass substrate (Corning Eagle 2000).  The
thickness of the silicon layer varies between 130 and 190~nm.
Figure~\ref{fig:raman} shows Raman measurements of liquid-phase precursor
sample \#1 and a PECVD reference sample.  Obviously, the peak at
480\,cm$^{-1}$, which is attributes to \mbox{a-Si:H}, is shifted to smaller
wavenumbers for the sample \#1.  As derived by
\cite{anastassakis1985physical}, the hydrostatic in-plane stress $\sigma$ and
change in Raman peak shift $\Delta\omega$ for single-crystal silicon are related
linearly:
\begin{equation}
  \label{eq:raman-shift-to-stress}
  \sigma = -249~\mathrm{MPa\,cm}\cdot\Delta\omega.
\end{equation}
For $\Delta\omega < 0$, this means $\sigma>0$, i.\,e.\ tensile stress.  In
accordance with \cite{ge2013detailed}, we apply
eq.~(\ref{eq:raman-shift-to-stress}) to the \mbox{a-Si:H} peak at 480\,nm, too.

After the fabrication, the samples were etched in hydrofluoric acid (20\,\%)
for approximatively 10~seconds.  Afterwards, they were carefully blow-dried
with nitrogen.

Figure~\ref{fig:crack} shows microscopic pictures of the samples after etching.
As you can see, spiral cracks have been formed.  In most cases, they are
interleaved double spirals but we also observed single ones.  The width~$w$ of
the spiral stripes is rather constant along the stripe length, and constant on
a given sample.  Furthermore, the two outer end points of the cracks are in
opposite position to each other, or very close to that position.  We observed
no prefered winding direction.  Table~\ref{tab:experimental-results} lists the
Raman peak shifts~$\Delta\omega$ and the spiral stripe widths~$w$ for the
measured samples.

\section{Model of crack formation}
\label{sec:model-crack-form}

\begin{figure}
  \centering
  \includegraphics[width=0.9\linewidth]{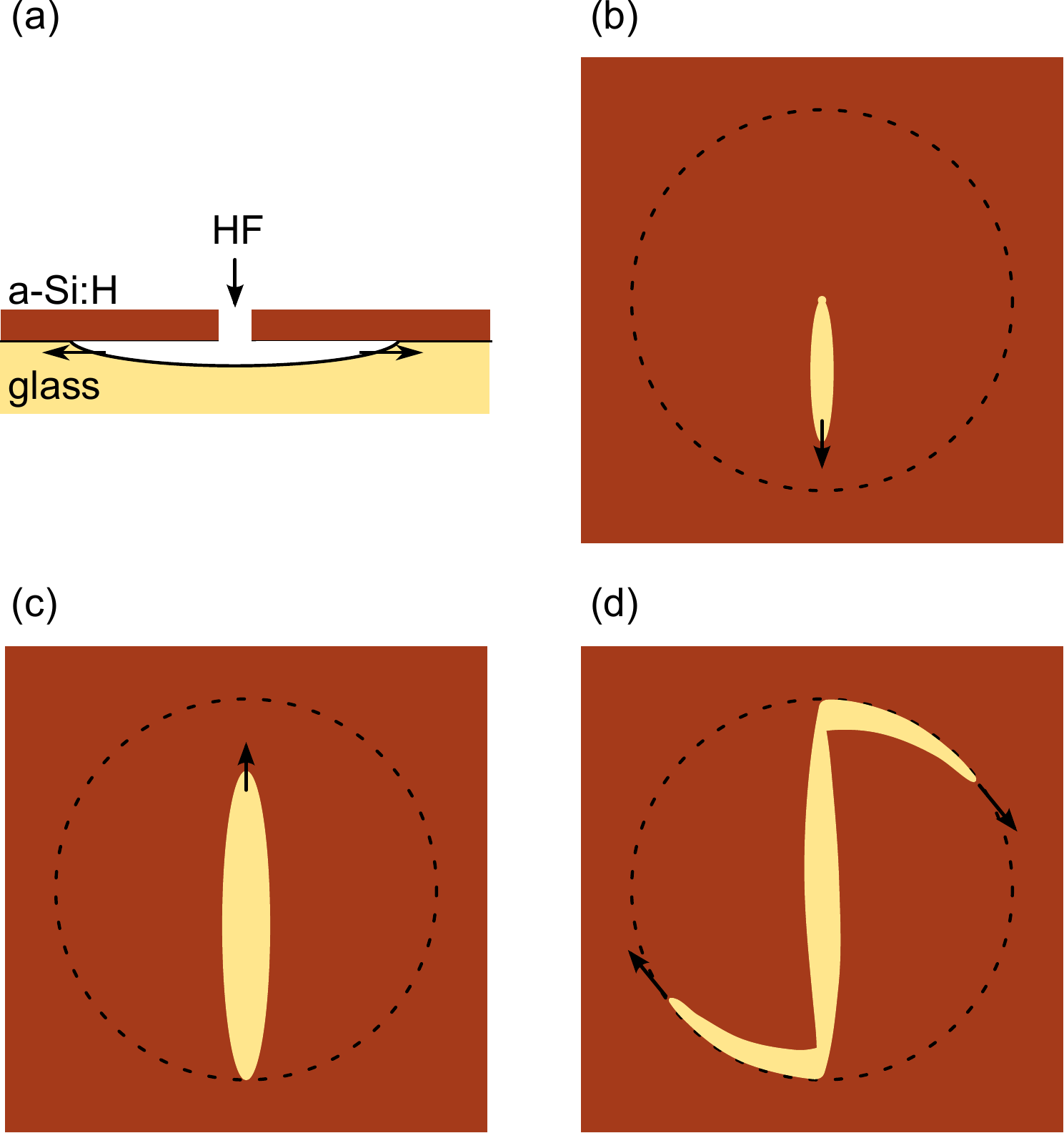}
  \caption{The crack propagation model for spiral fractures in \mbox{a-Si:H}\@.
    (a) is a cross section of the layer structure, (b--d) are top views.  The
    arrows denote the direction of etching (a) or crack propagation (b--d).
    The dashed circle is the area of underetching.  The colours mimic the
    colours in figure~\ref{fig:crack}.}
  \label{fig:crack-propagation}
\end{figure}

Figure~\ref{fig:crack-propagation} illustrates how we think these spiral cracks
come into existence during etching.

The hydrofluoric acid etches glass very efficiently.  In contrast, it leaves
silicon more or less unaffected.  The acid probably attacks a weakness in the
silicon layer, see figure~\ref{fig:crack-propagation}~(a).  This weakness may
be a tiny hole, or a small inclusion of etchable material, e.\,g.\ SiO$_2$.  In
any way, the acid penetrates the weakness and starts underetching the silicon.
This leads to a disk-shaped zone of underetching with the weakness in the
centre.  Above this zone, the silicon is unsupported and thus, the stress in
the material is no longer sustained by the substrate.  In
figure~\ref{fig:crack-propagation}~\mbox{(b--d)}, this zone is marked by the
dashed circle.

A small asymmetry of the initial hole, e.\,g.\ a notch in its edge,
concentrates the stress and begins to crack.  The crack tip propagates towards
the rim of the underetching, see figure~\ref{fig:crack-propagation}~(b).  While
doing so, the stress intensity factor increases according to
$\text{SIF}\sim\sqrt{\text{crack length}}$.  Note that this also drastically
increases the stress at the initial hole, so that eventually, a second crack
forms and grows towards the rim, see figure~\ref{fig:crack-propagation}~(c).

Both crack tips eventually reach the rim, one of them first.  This one
continues on the rim iteself, determining the winding direction of the final
spiral.  The slightly slower crack tip takes the same direction because the
shearing stress due to the asymmetrical situation caused by the already cracked
rim on the opposite side guides it whis way.

Both cracks continue on the rim until the relaxation has reduced the stress so
much that no further crack growth is possible, see
figure~\ref{fig:crack-propagation}~(d).  However, at the same time, the
underetched disk grows.  The newly detached silicon adds new stress to the
layer, which lets the cracks start growing again.  Both processes happen
simultaneously.  This way, the spirals are formed.  We assume that crack growth
is much faster than disk growth, so that the stress intensity factor (SIF) at
the crack tips always is very close to the facture toughness $K_{Ic}$ of the
material.  In particular, this means that the spiral stripe width is largely
independent of the kinematics of the processes, but only dependent on layer
properties.

\section{Estimation of fracture toughness}
\label{sec:estim-fract-toughn}

The only quantity easily accessible for measurement is the stripe width of the
spiral.  Therefore, it is important to understand the mechanism behind
it. Apparently, it is a constant for a certain sample, and in particular
independent of the distance from the centre of the spiral.  A plausible albeit
not rigorous explanation is the following:

\begin{figure*}
  \centering
  \includegraphics[height=0.4\linewidth]{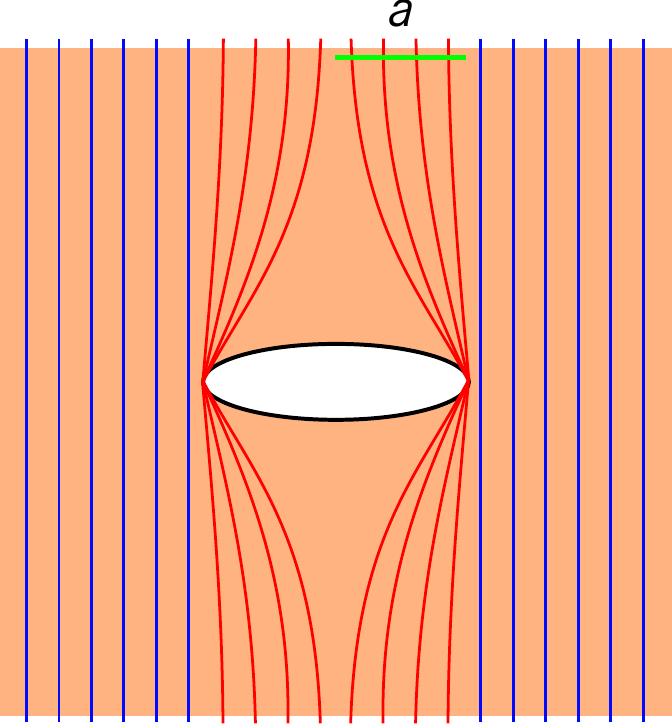}\hspace{1cm}%
  \includegraphics[height=0.4\linewidth]{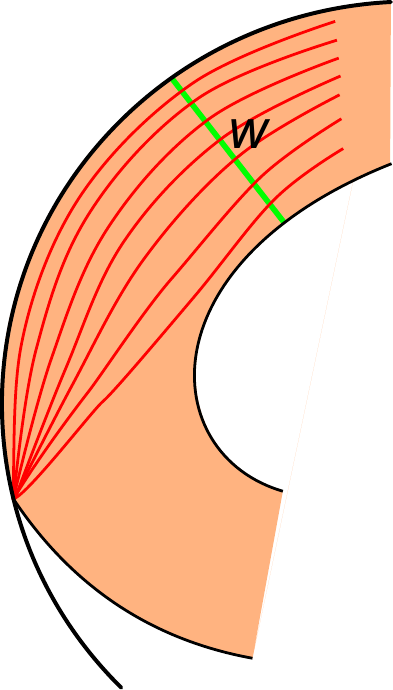}
  \caption{Comparison of the geometries of the Griffith fracture (left) with the
    situation in the spirals of the thin-film samples (right).  The
    characteristic crack length is marked in green.}
  \label{fig:griffith-analogy}
\end{figure*}

Figure~\ref{fig:griffith-analogy} compares the well-understood Griffith
fracture with our geometry.  The Griffith facture occurs in an infinitely large
workpiece which contains an initial crack.  Additionally, the workpiece is
stressed perpendicular to the crack.  Then, the stress intensity factor SIF is
given by
\begin{equation}\label{eq:1}
  \text{SIF} = \sigma\sqrt{\pi a} f,
\end{equation}
with $\sigma$ being the stress in the material, $a$ being the half crack
length, and $f$ being the geometry factor, which is unity for the Griffith
fracture.  $f$ deviate from unity if e.\,g.\ the angle between the crack and
the stress differs from 90\textdegree. It is important to see how the SIF is
generated: The lines of force from the length $a$ (shown in red) are deviated
to the crack tip because the edges of the crack are load-free.

The right-hand side of figure~\ref{fig:griffith-analogy} depicts the situation
in the spiral: The arc to the left is the border of the underetching.  The
crack tip is the join of the lower line coming from the right with the arc.
Since the inner (i.\,e.\ right) edge of the stripe is load-free, the radial
component of the stress vanishs. The circular component, however, is only
slightly weakend by the opening of the crack at the bottom.

The characteristic length in this case is the stripe width~$w$.  It corresponds
to the half crack length in the Griffith fracture because all lines of force
that end in the crack tip pass this width more or less
perpendicularly. Moreover, the circular stress decays only slightly towards the
inner edge, i.\,e.\ in can be assumed homogeneous like for the Griffith
fracture.

This suggests that eq.~(\ref{eq:1}) also applies here with $a=w$.  Finally,
assuming the model of crack formation of the previous section which implies
$\mathrm{SIF} = K_{Ic}$, this allows the estimation of facture toughness
according to
\begin{equation}
  \label{eq:fracture-toughness-griffith}
  K_{Ic} = \sigma\sqrt{\pi w} f.
\end{equation}

This result holds for single and double spirals alike, as long as the process
of crack formation, i.\,e.\ a slowly growing disk-shaped detachment, is the
same.  At least after the first winding of the spiral, the local geometry of
crack and forces is equivalent in both cases.  This is also what is observed
experimentally: For both types of spirals, $w$ is the same for a given sample.

The next task is to back this up by accurate simulation, which also helps to
determine the particular factor $f$ for the spiral geometry.

\section{FEM simulation}

Because the arguments presented in the previous section are non-rigorous, a
numerical simulation of the geometry was perfomed in order to get results on a
solid basis.  Additionally, such a simulation provides a value for the $f$
geometry factor.

We used the finite element method FEM for this numerical simulation.  The FEM
is a well-understood and reliable method in fracture mechanics.  Moreover,
there are plenty of programs and programming libraries that aid developing FEM
models.  In particular, we chose the SfePy programming library for our
research.  \cite{cimrman2008sfepy} It is free software, well-documented, and
mature.  Besides, the Python programming language allows for convenient
developing.

\begin{figure*}
  \centering
  \includegraphics[height=0.4\linewidth]{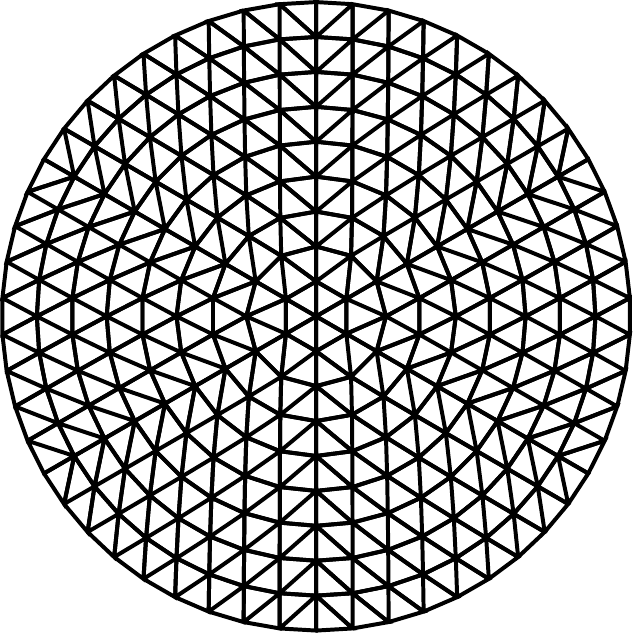}\hspace{1cm}%
  \includegraphics[height=0.4\linewidth]{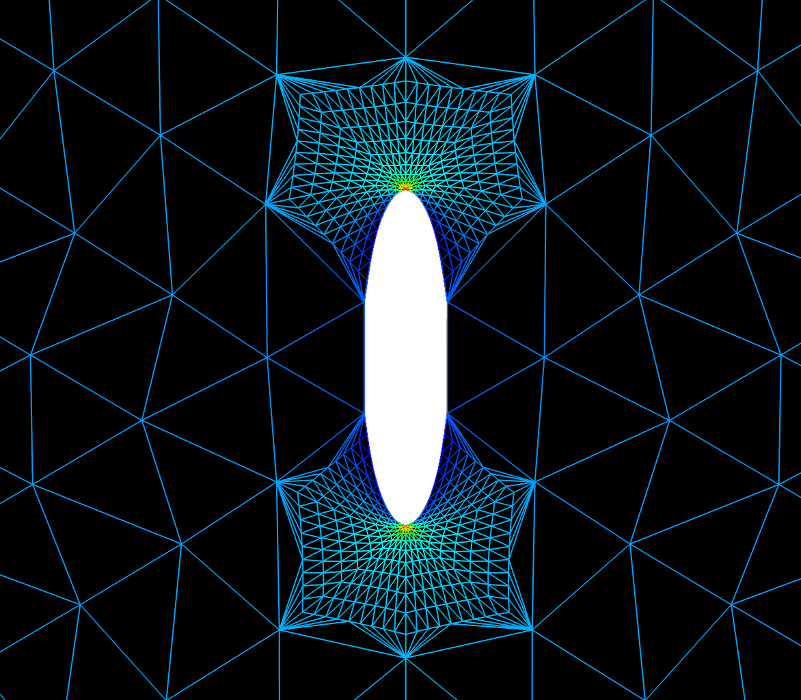}
  \caption{The triangulation used for the FEM simulations.  The triangular mesh
    for the disk (left), and the additional triangulation for the crack tips
    (right).  On the right-hand side, increasing tension is visualised by the
    colour sequence blue--cyan--green--yellow--red.}
  \label{fig:mesh}
\end{figure*}

We chose a triangular FEM mesh for covering the disk of underetched material,
see figure~\ref{fig:mesh}.  In this mesh, the current crack was realised by
severing the nodes along the crack path (duplicating the nodes).  Then, the
elements at the crack tips were replaced with hundreds of small sub-elements
which became even smaller towards the tip.  This way, we had the accuracy
needed for calculating reliable SIFs.

The SIFs were calculated with the ``displacement method''.  The displacements
of the nodes of the sub-elements (i.\,e.\ in a range of one big finite element)
were computed into SIF values which were extrapolated for $r\to0$, with $r$
being the distance to the tip.  We calculated both $K_I$ and $K_{\mathit{II}}$,
and used them for determining the equivalent SIF according to
\cite{richard2005theoretical} and the crack direction according to
\cite{alshoaibi2006finite}.

If the equivalent SIF exceeds the fracture toughness of the material, the crack
propagates.  Then, the crack direction determines which node is the next to be
severed.  This node is severed, and the new, longer crack has to be analysed by
FEM\@.  This is repeated until the material is relaxated enough so that the
crack cannot grow anymore.

It is difficult to simulate crack growth without any a priori assumptions
because then, you have to crack element after element, which uses a lot of
computation time.  Also, discretised directions impose inaccuracy to the
simulation: Since the polygonal elements have only a certain number of edges
originating at each vertex (node), the calculated crack direction has to be
constrained onto one of the edges.  This cannot be mitigated by increasing the
spatial resolution.

Therefore, we made some preliminary simulations to narrow the degrees of
freedom of the crack growth.  This showed the following:

\begin{enumerate}
\item Any initial crack, originating at the centre of the disk, grows straight
  towards the rim.
\item Any small deviation from the straight line makes the other tip of the
  crack deviate in the same direction (e.\,g.\ clockwise).  In other words, the
  tips try to evade each other.
\item When the crack propagates along the rim, its growth direction points
  outwards.  Thus, it must stay on the rim.
\end{enumerate}

Another trivial observation is that a symmetric initial crack (e.\,g.\ straight
through the centre over the total diameter) conserves symmetry over the whole
simulation.  This means that only one crack tip has to be analysed, and the
other one just mirrors its propagation.  These observations confirm the
formation model as presented in section~\ref{sec:model-crack-form}.

Hence, the outline of the simplified algorithm is as follows:

\begin{enumerate}
\item Create a straight initial crack through the total diameter of the initial
  disk.
\item Find the length of the crack on the rim of the disk for which the SIF at
  the tip equals the fracture toughness of the material, and let the crack grow
  to this length.\label{item:1}
\item Add another ring of finite elements around the disk.  This brings new
  stress into the material, and the crack can grow further.
\item If the winding number of the spiral is below a pre-set value, go back to
  (\ref{item:1}).  Otherwise, end the simulation.
\end{enumerate}

\begin{figure}
  \centering
  \includegraphics[width=0.9\linewidth]{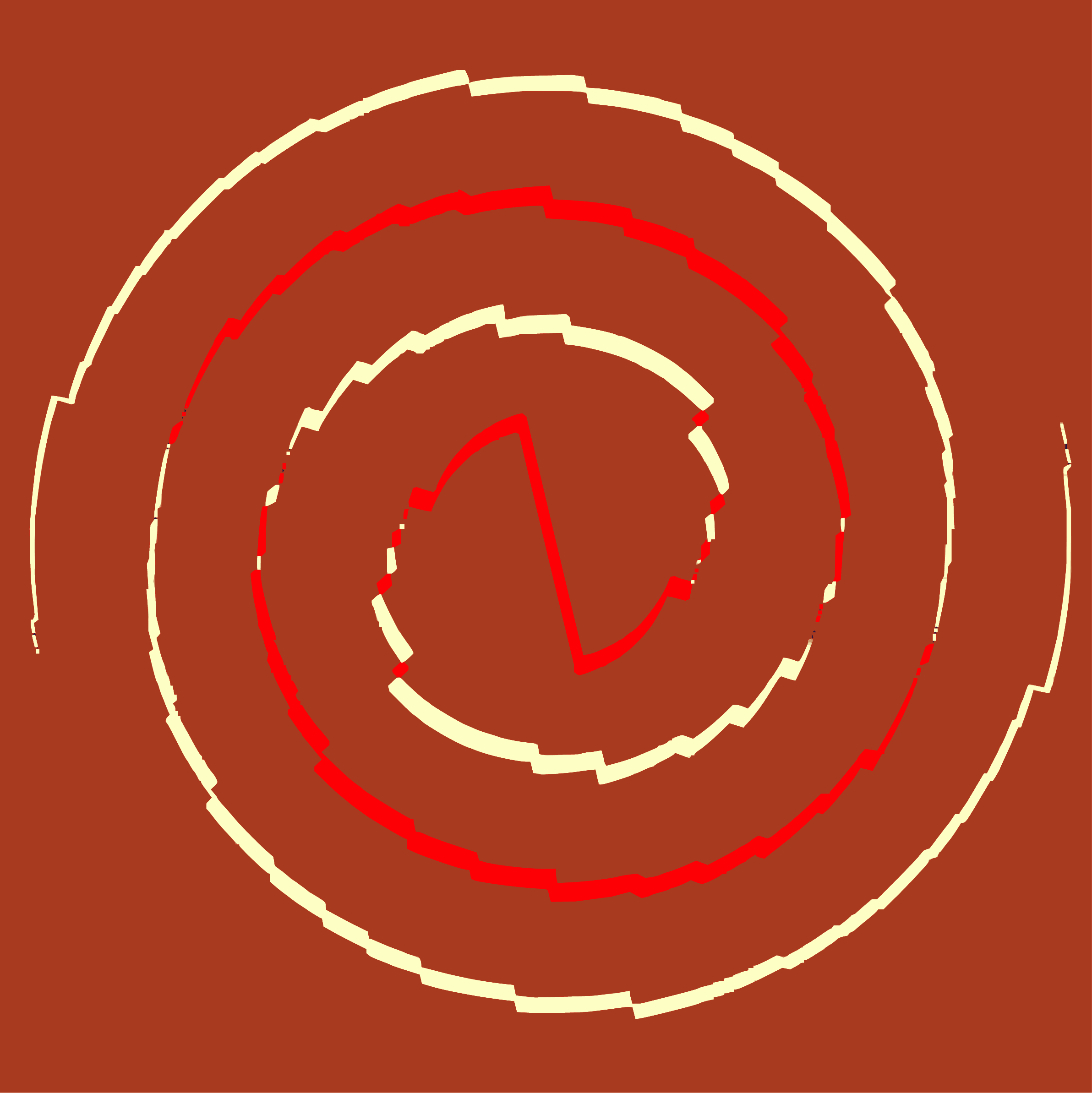}
  \caption{Visualisation of displacement as a result of an FEM simulation of
    crack growth in a layer with disk-shaped detachment from the substrate.
    You can see the spiral lobes being moved out of their original position due
    to stress forces.  The layer is brown, the substrate yellow, and two lobes
    of the layer stacked on top of each other are red.  These colours are
    chosen to match the colours of figure~\ref{fig:crack}.}
  \label{fig:fem-spiral}
\end{figure}

Figure~\ref{fig:fem-spiral} presents the result of the FEM in an illustrative
manner: The finite elements are displaced according to the values of tension in
the spiral lobes.  This way, it simulates a top view on the sample.  And
indeed, this image and the microscopic photograph in figure~\ref{fig:crack}
match very well.  The alternation of gap and overlay is properly reproduced, as
is the constant stripe width.

\begin{figure}
  \centering
  \includegraphics[width=0.9\linewidth]{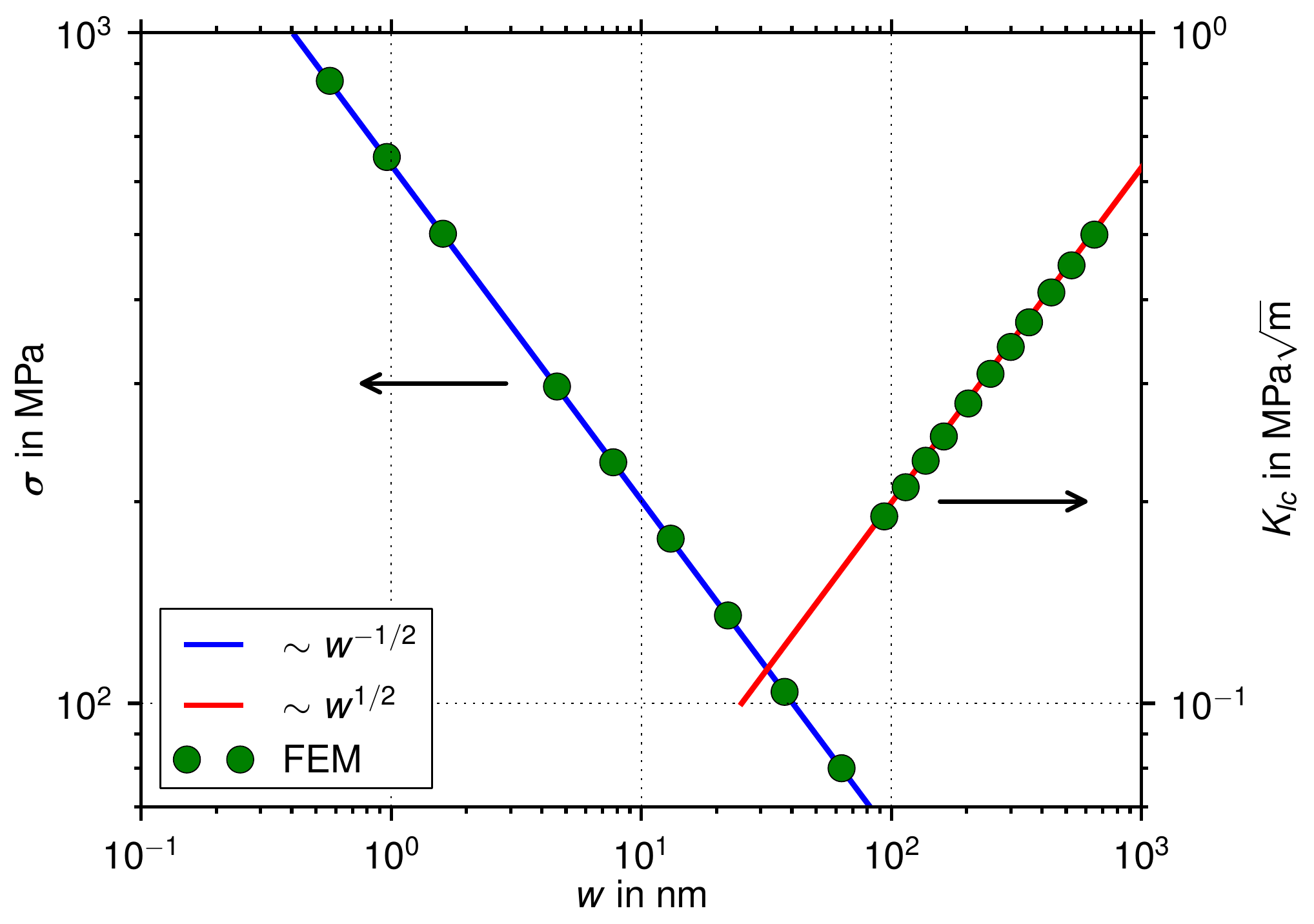}
  \caption{Disk tension $\sigma$ and fracture toughness $K_{Ic}$ for various
    spiral stripe widths $w$, as determined with FEM simulations.  The assumed
    fracture toughness for the $\sigma$ curve is $K_{Ic}=0.5~\mathrm{MPa\sqrt
      m}$.  The tension in the layer for the $K_{Ic}$ curve is 800~MPa.  The
    lines indicate a fit with eq.~(\ref{eq:fracture-toughness-griffith}) and
    $f=0.44$.}
  \label{fig:stripe_width}
\end{figure}

Figure~\ref{fig:stripe_width} analyses the FEM results quantitatively by
cutting two perpendicular planes out of the three-dimensional parameter space,
namely $\sigma(w)$ and $K_{Ic}(w)$.  The two lines correspond to
\begin{align}
  \label{eq:fem-tension}
  \sigma(w) &= \frac {K_{Ic}}{0.44\sqrt{\pi w}} \\
  K_{Ic}(w) &= 0.44\sigma \sqrt{\pi w}\;.
\end{align}
This equation set is compatible with
eq.~(\ref{eq:fracture-toughness-griffith}).  Thus, the FEM confirms the
analytical approach of section~\ref{sec:estim-fract-toughn} and yields a value
for the geometry factor:
\begin{equation}
  \label{eq:fem-final}
  K_{Ic}^{\text{spiral}} = \sigma\sqrt{\pi w} f,\quad f=0.44
\end{equation}

\section{Application to the experiments and discussion}

Applying eq.~(\ref{eq:raman-shift-to-stress}) to the results of
table~\ref{tab:experimental-results} and then using eq.~(\ref{eq:fem-final})
yields the facture toughness of each sample.  The averaged fracture toughness
of the amorphous silicon in the samples is
\begin{equation}
  \label{eq:toughness-a-si}
  K_{Ic}^{\text{a-Si:H}} = 4.7(3)~\mathrm{MPa\sqrt m}.
\end{equation}
This value should be compared with the $K_{Ic}$ of single-crystal silicon of
$0.82~\mathrm{MPa\sqrt m}$ \cite{mercado2003multichip} and that of
poly-crystalline silicon in the range 0.86--1.1~$\mathrm{MPa\sqrt m}$
\cite{chasiotis2006fracture,bagdahn2001fracture,kahn2000fracture}.
\cite{swadener2003increasing} reports a lower limit of $K_{Ic}$ of
\mbox{a-Si:H} at $1.35~\mathrm{MPa\sqrt m}$.  It has been frequently reported
(e.\,g.\ by \cite{swadener2003increasing,gilbert1997fracture}) that amorphous
phases exceed their crystalline counterparts in fracture toughness, sometimes
by more than a decade.  Thus, we consider the above $K_{Ic}^{\text{a-Si:H}}$ in
plausible accordance with previous findings.

We consider the variation of the measured stripe widths the dominating error in
the result.  Therefore, we used this as the standard error of the mean in
eq.~(\ref{eq:toughness-a-si}).  Other contributing errors are the difference
between \mbox{a-Si:H} and \mbox{c-Si} with respect to
eq.~(\ref{eq:raman-shift-to-stress}) and the uncertainty in the Raman peak
shift.

\section{Conclusions}

Thin layers of amorphous silicon from a liquid-phase polysilane precurser may
suffer from high residual tensile stress.  By underetching the layer through a
pinhole, cracks in form of single and double spirals are forming due to this
stress.  By making an analogy between the spiral cracks and the Griffith
fracture, one can identify the spiral stripe width with Griffith's crack
length.  An analysis with the FEM confirms this analogy and yields the geometry
factor for such spirals $f_{\text{spiral}}=0.44$.  The spiral stripe width
depends only on stress and fracture toughness, i.\,e.\ knowing one leads to the
other.  We determined the stress by Raman measurements, thus being able to
calculate the fracture toughness of the amorphous silicon
$K_{Ic}^{\text{a-Si:H}} = 4.7~\mathrm{MPa\sqrt m}$ with a standard error of
8\,\%.  This value exceeds that of crystalline silicon by a factor of~5.7.

Further investigation is necessary.  The independence of fracture toughness and
stress should be confirmed.  For this, samples covering a range of stress
intensities should be examined, and verified that $\sigma\sim1/\sqrt w$ for all
of them.  Moreover, eq.~(\ref{eq:raman-shift-to-stress}) should be calibrated
for \mbox{a-Si:H} using a direct method like a bendable substrate.  Finally,
pinholes may be created in the layer deliberately, trying to use the spiral
cracks as a very simple albeit destructive method to determine fracture
toughness for various types of thin-layer material.  This way, fracture
toughness may become an additional useful characterisation quantity for such
layers.

\bibliography{main}

\end{document}